# A non-volatile memory based on nonlinear magnetoelectric effects

Jianxin Shen[†], Junzhuang Cong[†], Yisheng Chai, Dashan Shang, Shipeng Shen, Kun Zhai, Ying Tian, and Young Sun[*]

Beijing National Laboratory for Condensed Matter Physics, Institute of Physics, Chinese Academy of Sciences, Beijing 100190, China

[†]These authors contributed equally to this work.
[*]Corresponding author. youngsun@iphy.ac.cn

Abstract

The magnetoelectric effects in multiferroics have a great potential in creating next-generation memory devices. We conceive a new concept of non-volatile memories based on a type of nonlinear magnetoelectric effects showing a butterfly-shaped hysteresis loop. The principle is to utilize the states of the magnetoelectric coefficient, instead of magnetization, electric polarization or resistance, to store binary information. Our experiments in a device made of the PMN-PT/Terfenol-D multiferroic heterostructure clearly demonstrate that the sign of the magnetoelectric coefficient can be repeatedly switched between positive and negative by applying electric fields, confirming the feasibility of this principle. This kind of non-volatile memory has outstanding practical virtues such as simple structure, easy operations in writing and reading, low power, fast speed, and diverse materials available.

The global information era has been seeking a universal non-volatile random-access memory (NVRAM) for many years. The best-known form of NVRAM today is flash memory. However, some drawbacks of flash memory prevent it from becoming the universal memory. Several competitive technologies are attempting to replace flash memory in certain roles. These include magnetic random access memory (MRAM) [1-3], resistive switching random access memory (RRAM) [4,5], phase change memory (PRAM) [6,7], ferroelectric random access memory (FeRAM) [8-10], and racetrack memory [11]. To date these alternatives of NVRAM have not yet become mainstream in industry because each of them faces certain challenging hurdles.

Multiferroics that combine magnetism and ferroelectricity as well as mutual coupling between them [12-15], hold a promise for designing a new generation of memory devices. Several strategies towards a non-volatile magnetoelectric (ME) random access memory or a multiferroic memory have been proposed and explored in the past decade [16-25]. One popular way is to integrate a ferroelectric layer in a magnetic tunnel junction to make a multiferroic tunnel junction [16,17,20]. The combination of the magnetoresistance (MR) and electroresistance (ER) effects could enable a four-state resistive memory device. Another general approach is to directly switch the magnetization of a multiferroic structure using an electric field for possible non-volatile devices [21-23]. Other policies such as combining the ME coupling with the interfacial exchange bias have also been studied [24,25]. Though conceptually appealing, these newly designed multiferroic memories with complicated structures face significant practical obstacles to be overcome before they could be commercialized.

In most of known NVRAMs, the binary information is generally stored by three quantities, respectively: the direction of magnetization, the direction of electric polarization, and the level of resistance. In this work, we propose that an alternative quantity – the sign of the ME coefficient, can be effectively used to encode binary information. The ME coefficient is defined as [12-14]: $\alpha_D = dP/dH$ and $\alpha_C = \mu_0 \, dM/dE$, where $P$ is electric polarization, $M$ is magnetization, $H$ is magnetic field, and $E$ is electric field. The former is called the direct ME effect and the latter is called the converse ME effect. Both $\alpha_D$ and $\alpha_C$ can be either positive or negative, depending on the ME coupling mechanism and process.

As shown in Fig. 1(a), the ME effects of a multiferroic system can be plotted by the relationship between $P$ and $H$ or $M$ and $E$. In general, the ME effects are classified into the linear and higher-order (nonlinear) effects [12]. The nonlinear ME effects may exhibit various $P$–$H$ or $M$–$E$ relationship deviating from linearity. In certain cases, a butterfly-shaped hysteresis loop could appear when the applied $H$ or $E$ field exceeds a critical field. This critical field is typically the coercive field that reverses the direction of $M$ or $P$. In fact, such kind of butterfly-shaped hysteresis has been widely observed in multiferroic heterostructures made of magnetostrictive and piezoelectric components [26-28]. One characteristic of the butterfly-shaped loop is that its two branches have opposite signs of slope. Thus, the ME coefficient ($\alpha_D$ or $\alpha_C$) has two distinctive states, positive and negative, corresponding to binary information "0" and "1", respectively. In the low-field region, the sign of the ME coefficient does not change with external fields. Only after a high field above the critical field is applied, it inverts its sign and retains the last state.

Based on this principle, a non-volatile memory element can be built up, shown in Fig. 1(b). It is a simple sandwich structure consisting of a multiferroic medium and two electrodes. For the benefit of easy operations in write and read, we consider a multiferroic (either single-phase or composite) with in-plane $M$ and vertical $P$. This multiferroic medium holds a nonlinear ME effect with a butterfly-shaped $M$-$E$ relationship similar to Fig. 1(a), which allows the binary information – the sign of $\alpha_C$ is written electrically by applying a voltage between two electrodes. To read out information, one normally needs to measure the ME coefficient $\alpha_C$ by applying a low $E$ field and detecting the induced small change of $M$. However, this reading mode will complicate the device structure because additional layers such as a magnetic tunneling junction is required to detect the change of local magnetization. Fortunately, this problem can be solved by measuring $\alpha_D$ instead. In the low-field regime, the ME effects are almost linear so that $\alpha_D$ and $\alpha_C$ are nearly equal; at least, they must have the same sign. Therefore, to read information, one can simply measure $\alpha_D$ rather than $\alpha_C$. The measurement of $\alpha_D$ is much easier, by applying a low $H$ and detecting the induced change of $P$. In practice, the ME voltage coefficient $\alpha_E = dE/dH \propto \alpha_D$, is measured instead, by applying a low magnetic field ($\Delta H$) and detecting the induced voltage ($\Delta V$) between two electrodes (Fig. 1(c)) – a technique that has been widely used in the study of ME composites [12-14]. In this way, both the writing and reading operations are convenient and efficient.

To testify the feasibility of the above principle, we have performed experiments in a device made of PMN-PT(110)/Terfenol-D heterostructure. PMN-PT is a well-known ferroelectric with a large piezoelectric effect and Terfenol-D ($Tb_{0.28}Dy_{0.72}Fe_{1.95}$) is a famous magnetostrictive material. Thus, this typical multiferroic heterostructure has pronounced ME coupling via the interfacial strain. Figure 2(a) shows the structure of the device and the configuration of experiments. The $E$ field is applied between two Ag electrodes to switch the direction of $P$ in PMN-PT layer and the dc magnetic field $H_{dc}$ is applied in plane to alter the $M$ of Terfenol-D layer. For the measurement of $\alpha_E$, a conventional dynamic technique is used.

We first checked the $M$–$E$ relationship of this device. As shown in Fig. 2(b), the in-plane $M$ varies with applied vertical $E$, and a butterfly-shaped hysteresis loop indeed appears as expected. The turning points of $M$ lie around ± 2 kVcm$^{-1}$, which match with the coercive fields of the ferroelectric PMN-PT layer, as deduced from the $I$-$E$ curves shown in Fig. 2C. This result confirms the basis of our principle for memory: upon the reversal of $P$, the slope of the $M$–$E$ relationship and consequently the sign of $\alpha_C$ invert.

Figure 3 shows how the ME voltage coefficient $\alpha_E$ of the device depends on the status of both $M$ and $P$. Before measuring $\alpha_E$, the device was pre-poled by applying a positive or a negative $E$ field of 4 kVcm$^{-1}$ to set the direction of $P$. Then, $\alpha_E$ was measured as a function of in-plane $H_{dc}$. When $P$ is set to upward (the red curve), $\alpha_E$ is very small in the high $H$ region because $M$ is saturated and the magnetostriction coefficient is nearly zero. As $H$ decreases from 10 kOe to zero, $\alpha_E$ increases steadily and exhibits a maximum ($\approx$ 160 mVcm$^{-1}$Oe$^{-1}$) around 1 kOe where the magnetostriction coefficient of Terfenol-D reaches a peak. When $H$ scans from positive to negative, $\alpha_E$ also changes its sign from positive to negative and shows a minimum ($\approx$ –160 mVcm$^{-1}$Oe$^{-1}$) around –1 kOe. In contrast, when $P$ is set to downward (the black curve), the $H$ dependence of $\alpha_E$ is totally opposite, being negative for +$H$ and positive for –$H$. These results reveals that the sign of $\alpha_E$ depends on the relative orientation between $M$ and $P$: for fixed direction of $P$, the sign of $\alpha_E$ can be switched by reversing $M$ with $H$; for fixed direction of $M$, the sign of $\alpha_E$ can be switched by reversing $P$ with $E$. The latter case is employed for the non-volatile memory in this work. We note that there is a small hysteresis in the $H$ dependence of $\alpha_E$, which means that $\alpha_E$ does not drop to null in zero $H$. This would be favorable to practical applications as we will discuss later.

For a memory element, the binary information has to be repeatedly written and read for many times. Figure 4 demonstrates the repeatable switch of $\alpha_E$ of the device. After applying a +4 kVcm$^{-1}$ $E$, $\alpha_E$ was measured for ~ 100 seconds; then, a −4 kVcm$^{-1}$ $E$ was applied to reverse $P$ and $\alpha_E$ was measured for another ~ 100 seconds. This process was repeated for six cycles. Figure 4(a) shows the experimental results measured with a dc bias $H_{dc}$=1 kOe where $\alpha_E$ is about maximum. Just as we expected, once the applied $E$ reverses $P$, $\alpha_E$ inverts its sign and retains the state until next application of $E$. For practical applications, it is better to operate without a dc bias $H_{dc}$. Thanks to the hysteresis shown in Fig. 3, $\alpha_E$ does not drop to null at zero $H$. Figure 4(b) shows the repeatable switch of $\alpha_E$ measured without a dc bias $H_{dc}$. Though $\alpha_E$ decreases from ~160 mVcm$^{-1}$Oe$^{-1}$ in 1 kOe to ~ 7 mVcm$^{-1}$Oe$^{-1}$ in zero $H_{dc}$, the repeatable switch of $\alpha_E$ with periodical $E$ is still clearly seen.

The above experiments confirm the feasibility of our new principle of a non-volatile memory: the sign of the ME coefficient can be encoded as binary information. This new policy is in contrast to other NVRAMs where the direction of $M$ or $P$ and the level of resistance are usually used for binary information. This kind of memory based on the ME coefficient has outstanding advantages. Firstly, the memory element has a very simple sandwich structure so that it is easy to fabricate and scale down for high-density storage. Secondly, the writing operation is easy and fast, by applying a voltage pulse between two electrodes, like that in a FeRAM and a RRAM. Thirdly, the reading operation now becomes convenient because it avoids the destructive read of $P$ in a FeRAM and the inconvenient read of local $M$ in a MRAM. The information is read out by simply measuring the sign of voltage across the electrodes while an independent coil supplies a small $H$. In fact, as illustrated in Fig. 1(c), all the memory elements can share a single reading coil and all the stored information can be read out in a parallel way. This would further simplify the structure and operations of the whole memory device. Fourthly, as the device is made of insulating multiferroics and both the write and read operations avoid considerable currents, it has a very low power consumption. Finally, a large amount of materials are available to make the memory device. Here, as the first demo of principle, we used complex compounds like Terfenol-D and PMN-PT due to their superior magnetostrictive and piezoelectric properties. In principle, many ferromagnet/ferroelectric heterostructures can act as the memory element as long as their ME coefficients can be reliably detected. Of course, to optimize the performance of the memory device for industry applications, a careful selection of materials and design of size and structures need to be systematically investigated in the future.

Lastly, we want to point out that the memory element, *i.e.*, the device shown in Fig. 1(b), has a more fundamental meaning in science: it is actually the fourth memelement, in addition to memristor, memcapacitor, and meminductor. As we discussed in our recent theoretical paper [29], the device of Fig. 1(b) holds both charge $q$ and magnetic flux $\varphi$ and yields a $\varphi-q$ relationship via the direct and converse ME effects, respectively:

$$dq = g\frac{\alpha_D}{\mu_0\mu_r}d\varphi \tag{1}$$

$$d\varphi = g\frac{\alpha_C}{\varepsilon_0\varepsilon_r}dq \tag{2}$$

where g is a geometric factor, $\varepsilon_0$ ($\mu_0$) are the permittivity (permeability) of vacuum and $\varepsilon_r$ ($\mu_r$) are the relative permittivity (permeability) of the ME medium, respectively. Consequently, a new quantity called transtance, $T = d\varphi/dq$, can be defined for both cases. Transtance is parallel to resistance (defined from $i$-$v$ relationship), capacitance (defined from $q$-$v$ relationship), and inductance (defined from $i$-$\varphi$ relationship). When the device shows a linear $\varphi$–$q$ relationship, it is a linear transtor – the fourth fundamental circuit element in addition to linear resistor, capacitor, and inductor. When the device exhibits a pinched or butterfly-shaped hysteresis loop, it is a nonlinear memtranstor. Just like the memristor that is considered as the element for a RRAM, the memtranstor has been predicted by us to function as a memory element as well, a conjecture verified in this work. Since transtance ($T$) has a more fundamental significance in electric circuits than the ME coefficients $\alpha_D$ and $\alpha_C$, we intend to term this new type of non-volatile memory based on nonlinear ME effects as a transtance switching random access memory (TRAM). TRAM opens up an exciting new door towards a universal memory.

**Acknowledgments**
This work was supported by the National Natural Science Foundation of China (Grant Nos. 11227405, 11534015, 51371192, 11274363), and the Chinese Academy of Sciences (Grants No. XDB07030200 and KJZD-EW-M05).

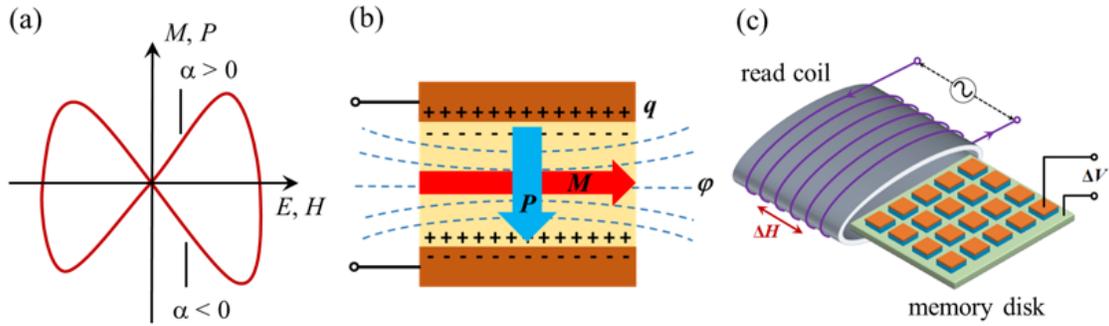

Fig. 1. (a) The butterfly-shaped hysteresis of the *M-E* or *P-H* relationship of multiferroics due to the nonlinear ME effects. The two states (positive and negative) of the ME coefficient α can be used to store binary information. (b) The schematic structure of a memory element. It consists of a multiferroic medium with in-plane magnetization (*M*) and out-of-plane electric polarization (*P*) sandwiched between two electrodes. (c) The illustration of read operation. The array of memory elements is put into a read coil that generates a small magnetic field $\Delta H$. The stored binary information (the sign of α) is read out by measuring the sign of induced voltage $\Delta V$. All the memory elements can share a single read coil, which greatly simplifies the fabrication and operations of the memory device.

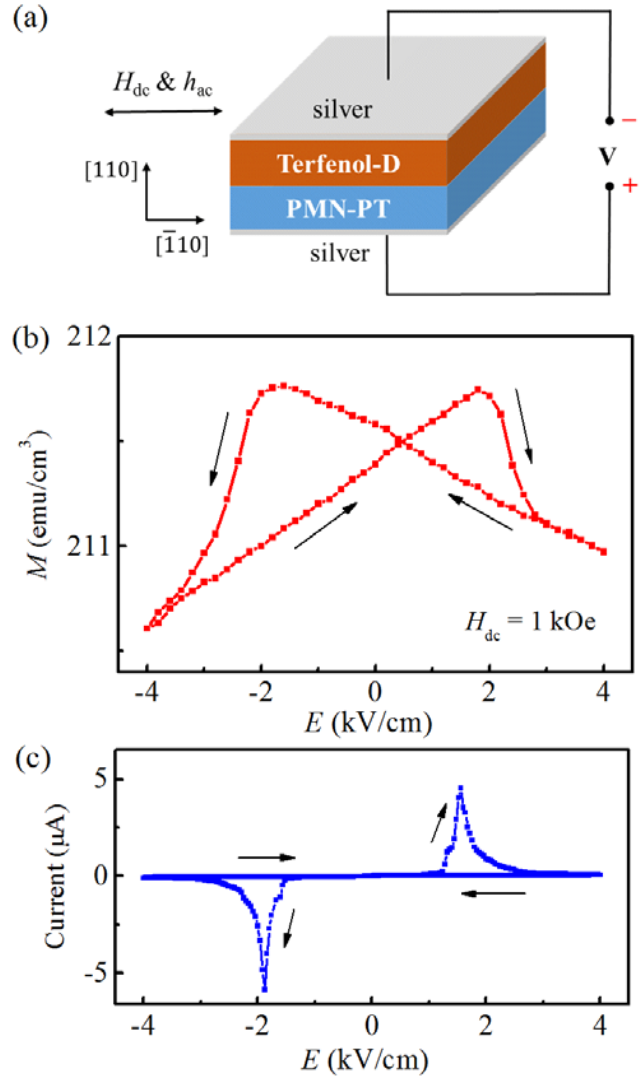

Fig. 2. (a) The structure of the device and the measurement configuration. The *E* field is applied vertically along [110] of PMN-PT and both the dc bias and ac *H* fields are applied in plane along [-110] of PMN-PT. (b) The in-plane magnetization of the device as a function of electric field. A butterfly-shaped hysteresis is observed. (c) The current-voltage curve of the device. The current peaks reflect the reversal of *P* in PMN-PT layer at coercive fields $\approx \pm 2$ kVcm$^{-1}$.

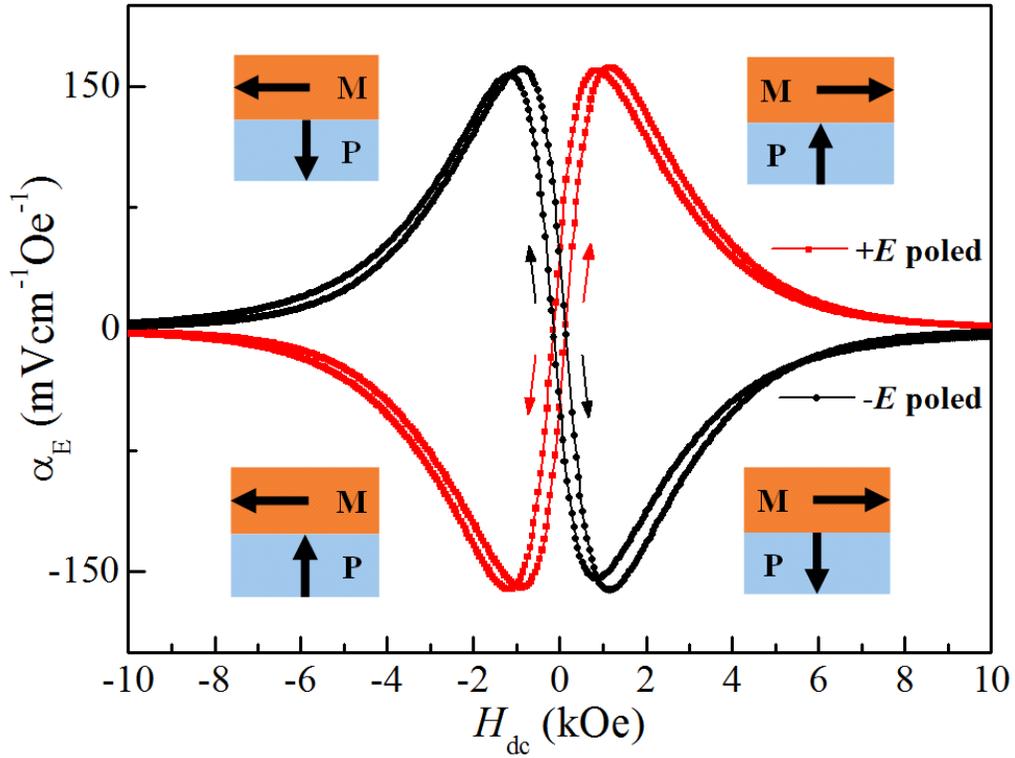

Fig. 3. The ME voltage coefficient $\alpha_E$ of the device as a function of dc bias magnetic field. The state of $\alpha_E$ depends on the relative orientation between $M$ and $P$. When the direction of $P$ is fixed, the reversal of $M$ by external magnetic fields results in the sign change of $\alpha_E$. Alternatively, when the direction of $M$ is fixed, $\alpha_E$ also changes its sign upon the reversal of $P$ by external electric fields. The latter case is used for the non-volatile memory in this work.

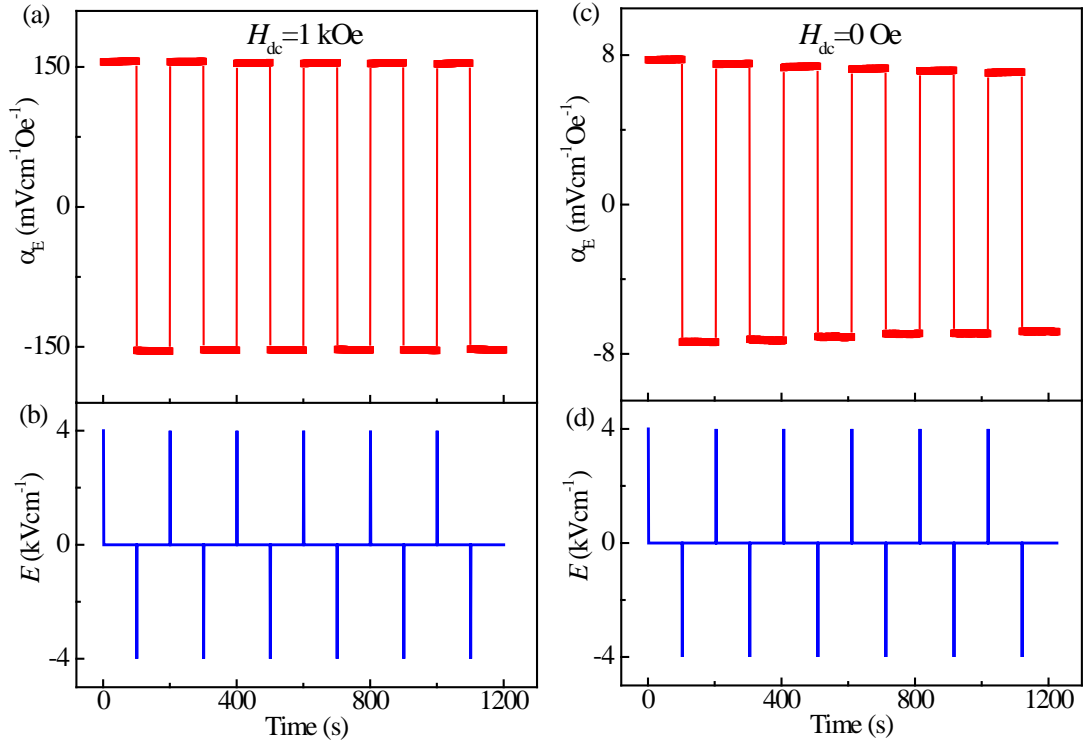

Fig. 4. Repeatable switch of $\alpha_E$ by applying electric fields. $\alpha_E$ of the device as a function of time measured under (a) 1 kOe dc bias magnetic field and (c) zero dc bias field. (b) and (b) The applied electric field as a function of time.